\documentclass[12pt,a4paper]{iopart}
\pdfoutput=1
\usepackage{graphics}
\usepackage{iopams} 
\usepackage{cite}

\newcommand{\be}{\begin{equation}}
\newcommand{\ee}{\end{equation}}

\newcommand{\ra}{\rightarrow}
\newcommand{\D}{\mathrm{d}}

\newcommand{\E}{\mathrm{e}}
\newcommand{\mL}{\mathcal{L}}
\newcommand{\mH}{\mathcal{H}}

\newcommand{\reals}{\mathbb{R}}
\newcommand{\p}{\partial}

\begin{document}

\title[Non-classical large deviations for a noisy system with non-isolated attractors]{Non-classical large deviations for a noisy system with non-isolated attractors}

\author{Freddy Bouchet}
\address{Laboratoire de Physique, ENS Lyon, 46 all\'ee d'Italie, 69364 Lyon, France}

\author{Hugo Touchette}
\address{School of Mathematical Sciences, Queen Mary University of London, London E1 4NS, UK}

\date{\today}

\begin{abstract}
We study the large deviations of a simple noise-perturbed dynamical system having continuous sets of steady states, which mimick those found in some partial differential equations related, for example, to turbulence problems. The system is a two-dimensional nonlinear Langevin equation involving a dissipative, non-potential force, which has the essential effect of creating a line of stable fixed points (attracting line) touching a line of unstable fixed points (repelling line). Using different analytical and numerical techniques, we show that the stationary distribution of this system satisfies, in the low-noise limit, a large deviation principle containing two competing terms: i) a ``classical'' but sub-dominant large deviation term, which can be derived from the Freidlin-Wentzell theory of large deviations by studying the fluctuation paths or \emph{instantons} of the system near the attracting line, and ii) a dominant large deviation term, which does not follow from the Freidlin-Wentzell theory, as it is related to fluctuation paths of zero action, referred to as \emph{sub-instantons}, emanating from the repelling line. We discuss the nature of these sub-instantons, and show how they arise from the connection between the attracting and repelling lines. We also discuss in a more general way how we expect these to arise in more general stochastic systems having connected sets of stable and unstable fixed points, and how they should determine the large deviation properties of these systems.
\end{abstract}



\section{Introduction}
\label{secintro}

The dynamics of physical systems described by partial differential equations, such as those appearing in hydrodynamics, optics or quantum physics, are in many cases qualitatively similar to that of finite-dimensional dynamical systems, especially when an important dissipation mechanism is involved. When this is the case, instabilities, bifurcations, limit cycles and attractors are indeed often similar to their finite-dimensional counterparts  (see, e.g.,~\cite{guckenheimer1983}). By contrast, when dissipative mechanism do not exist or are very small in partial differential equations, these may exhibit  phenomena that have no counterparts in finite-dimensional systems. Examples of such phenomena include non-dissipative relaxation and asymptotic stability \cite{Pego_Weinstein_1994CMaPh.164..305P,Bouchet_Morita_2010PhyD,Mouhot_Villani:2009}, solitons, as well as the appearance of an infinite number of conserved quantities and infinite number of steady states \cite{Holm_etal_PhysRep_1985,Morrison_1998_HamiltonianFluid_RvMP,Arnold_1966,BouchetVenaille-PhysicsReport}.

In this work, we are interested in studying the rare events or large deviations \cite{touchette2009} of systems possessing continuous sets of steady states which effectively act as attractors. Such sets of steady states are found in many dynamical equations having a non-canonical Hamiltonian structure, such as the 2D Euler equation, the Vlasov equation, magneto-hydrodynamic equations, and the shallow-water equations, to mention only a few examples; see \cite{Holm_etal_PhysRep_1985,BouchetVenaille-PhysicsReport} for more. In all of these examples, the consequence of the non-canonical structure is the existence of an infinite number of conserved quantities or Casimirs, which are responsible for the infinite (and continuous) set of steady states \cite{Holm_etal_PhysRep_1985,BouchetVenaille-PhysicsReport}. Physically, these states are important because they can act as attractors, as has been found in experiments \cite{Sommeria_2001_CoursLesHouches,Tabeling02} and numerical simulations \cite{Schneider_Farge_2008PhysicaD}. Moreover, in some cases, their attractive behavior can be explained theoretically \cite{Mouhot_Villani:2009,Bouchet_Morita_2010PhyD} using arguments and methods based on statistical mechanics \cite{SommeriaRobert:1991_JFM_meca_Stat,Miller:1990_PRL_Meca_Stat,Bouchet_Sommeria:2002_JFM,BouchetVenaille-PhysicsReport}.

The problem that we address here is how the existence of a continuous set of steady states of a system influences its large deviation properties and how these properties compare with those of finite-dimensional systems. For the latter systems, two generic classes of systems have been considered from the large deviation point of view, namely:~i)~strongly dissipative systems with well-defined and disconnected attractors, as exemplified by the gradient dynamics of a Brownian particle in a potential (Kramers problem) \cite{kampen1992}, and~ii)~weakly-perturbed Hamiltonian systems with added weak friction and noise. The large deviations of both of these classes are well known: they can be obtained using semi-classical (WKB or instanton) approximations of path integrals \cite{graham1989,moss19891,mannella1990,dykman1994b,luchinsky1997,luchinsky1998,touchette2009} or, more rigorously, using the theory developed by Freidlin and Wentzell \cite{freidlin1984}.

Dynamical systems having an infinite number of steady states do not strictly fall in either of these generic classes. The fact that the dynamics of these systems is effectively irreversible and converge towards attractors that are asymptotically stable suggests that they are analogous to the class of strongly dissipative systems. However, because their steady states form one or several connected sets, the presence of a weak noise should lead these systems to diffuse over their attractors, as is the case for Hamiltonian systems. From this point of view, systems with infinite steady states share aspects of both strongly dissipative systems and weakly-perturbed Hamiltonian systems.

To illustrate this point, we consider in this paper a simple model with two degrees of freedom, denoted by $A$ and $B$,  having a continuous set of steady states in the zero-noise limit. Our goal in the following is to obtain the stationary probability density $P(A,B)$ of this model, which we refer to simply as the $AB$ model, in the low-noise limit $\nu\ra 0$, where $\sqrt{\nu}$ is the noise amplitude. Following the theory of large deviations, we expect this density to have the approximate form
\be
P(A,B)\approx\E^{-c_\nu I(A,B)},
\label{eqldt1}
\ee
in the limit $\nu\ra 0$, where $c_{\nu}$ is a coefficient, called the \emph{speed}, which diverges as $\nu\ra0$, and $I(A,B)$ is a $\nu$-independent function, called the \emph{rate function} or \emph{quasi-potential}. Large deviation approximations of this form have been extensively studied, as mentioned above, for various noise-perturbed dynamical systems, including systems with single-point attractors as well as systems with attracting limit cycles (see, e.g., \cite{graham1989,moss19891,mannella1990,dykman1994b,luchinsky1997,luchinsky1998}). In nearly all the classical examples that we are aware of, the speed of the large deviation approximation is $1/\nu$. In the case, for example, of a Langevin dynamics describing the overdamped motion of a Brownian particle in a potential $V(x)$ and in a fluid at temperature $T$, the speed is $1/(k_{B}T)$, with $k_{B}$ the Boltzmann constant, while the rate function is the potential $V(x)$. The resulting large deviation form for the stationary density,
\be
P(x)\approx\E^{-V(x)/(k_BT)},
\ee 
is in agreement with the known Arrhenius factor describing transitions probability between stable or metastable states (Kramers theory), and implies, as physically expected, that $P(x)$ concentrates in the low-noise limit (or low-temperature limit) on the equilibrium state minimizing the potential.

The results that we obtain for the $AB$ model show that $P(A,B)$ concentrates in a similar way on the set of stable steady states of the model -- in this case, a whole line of steady states -- but does so, in contrast with the classical cases, with a speed proportional to $1/\sqrt{\nu}$. Moreover, we show that close to the line of stable steady states, there is a correction to this large deviation approximation having a speed proportional to $1/\nu$. For a small but finite noise power $\nu$, $P(A,B)$ is therefore the sum of two contributions: a large deviation approximation with speed $1/\sqrt{\nu}$, which accurately describes the form of $P(A,B)$ away from the attractor, and a large deviation approximation with speed $1/\nu$, which describes $P(A,B)$ close to the attractor. 

These competing large deviation terms are related in the $AB$ model to different types of fluctuation paths having different large deviation speeds and, notably, to fluctuation paths of zero action, which we call ``sub-instantons''.  They are not related, we should mention, to the divergence of sub-leading prefactors entering in the large deviation approximation, discussed, e.g., by Berglund and Gentz \cite{berglund2009,berglund2010}. We argue in the concluding section of the paper that these sub-instantons should arise in more general systems having multiple connected sets of steady states, and that they should lead, as in the $AB$ model, to stationary probability densities having competing large deviation terms in the low-noise limit. 

The rest of the paper is organized as follow. In Sec.~\ref{secmodel}, we introduce the $AB$ model, and then proceed in Secs.~\ref{secpi}-\ref{sechj} to obtain $P(A,B)$ using three different approaches: a path integral approach which leads to results similar to those obtained in the framework of the Freidlin-Wentzell theory (Sec.~\ref{secpi}), a dynamical approach based on certain approximations of the $AB$ model (Sec.~\ref{secda}), and an approach based on the Hamilton-Jacobi equation (Sec.~\ref{sechj}). Throughout these sections, the analytical results obtained are compared with numerical results obtained by solving the Fokker-Planck equation directly. We conclude in Sec.~\ref{secconc} with some remarks on the generality of our results.

\section{Model}
\label{secmodel}

The model that we study is defined by the following set of two coupled (It\^o) stochastic differential equations (SDEs):
\begin{eqnarray}
\D A &=  (-AB-\nu A)\,\D t+\sigma_{A}\sqrt{\nu}\,\D W_{A}\nonumber\\
\D B &=  (A^{2}-\nu B)\,\D t+\sigma_{B}\sqrt{\nu}\,\D W_{B},
\label{eqAB1}
\end{eqnarray}
where $\nu>0$ is a real coefficient that balances, together with the positive constants $\sigma_A$ and $\sigma_B$, the dissipation of the model and the intensity of the two uncorrelated Brownian motions $W_A(t)$ and $W_B(t)$. Henceforth, we refer to this model simply as the $AB$ model. Note that the inclusion of the two dissipative or friction forces $-\nu A$ and $-\nu B$ prevents the system to escape (by diffusion) to $B=\infty$ as $t\ra\infty$, and makes sure, therefore, that a stationary density $P(A,B)$ exists.

\begin{figure}[t]
\centering
\resizebox{2.2in}{!}{\includegraphics{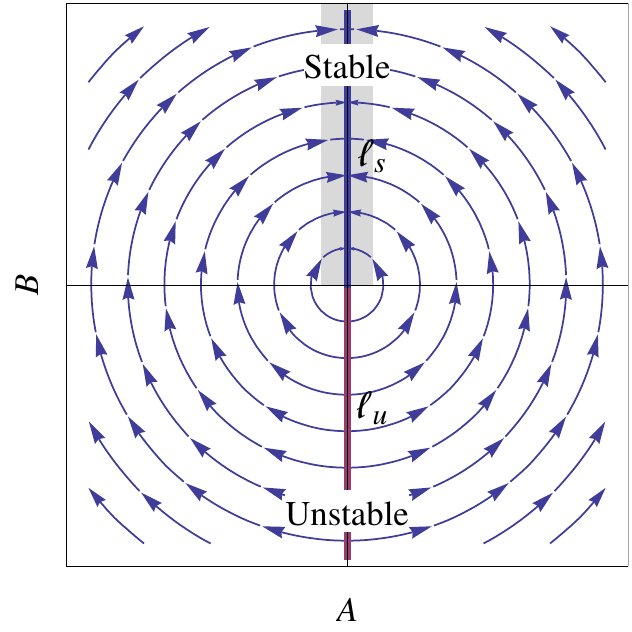}}
\caption{Vector field of the $AB$ model giving rise to a line $\ell_s$ of stable fixed points and a line $\ell_u$ of unstable fixed points.}
\label{figvf1}
\end{figure}

The zero-noise and zero-friction dynamics of the $AB$ model is given by
\begin{eqnarray}
\dot A&=-A B\nonumber\\
\dot B&=A^2.
\label{eqdets1}
\end{eqnarray}
This simple dynamics has the key properties that we referred to in the introduction, namely, 
\begin{enumerate}
\item It has a continuous set of steady states, which corresponds here to the line $A=0$;
\item The dynamics is irreversible and converges to an attractor, corresponding here to the upper semi-line $A=0$, $B>0$, which we denote by $\ell_s$. This semi-line is a line of \emph{stable} steady states or \emph{stable} fixed-points, as shown in Fig.~\ref{figvf1}. The lower semi-line $A=0$, $B<0$, denoted by $\ell_u$, is a line of \emph{unstable} fixed-points.\footnote{The connecting point $(0,0)$ is marginally stable.} 
\item The $AB$ dynamics conserves the quantity $E=A^2+B^2$, which we refer to as the energy.
\end{enumerate}

\begin{figure*}[t]
\resizebox{\textwidth}{!}{\includegraphics{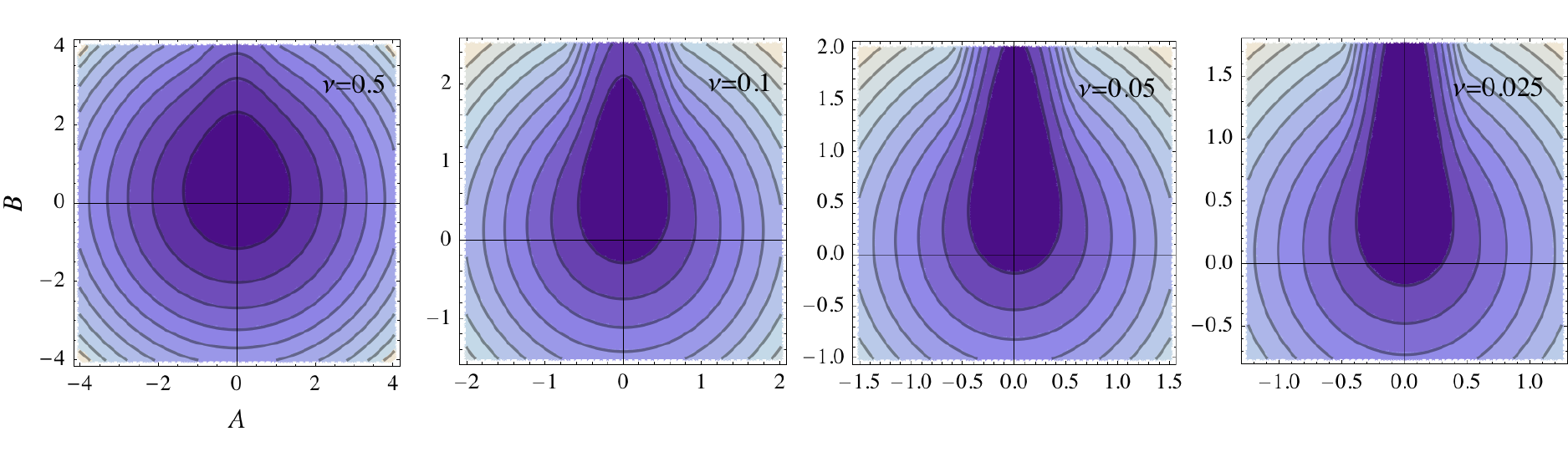}}
\caption{Contour plots of $P(A,B)$ obtained by direct numerical integration of the Fokker-Planck equation associated with the $AB$ model for $\sigma_A=\sigma_B=1$. The noise intensity $\nu$ used in the integration is shown in the plots. Darker colors correspond to larger values of $P(A,B)$. The region plotted for each noise power is different to highlight the concentration around the stable line.}
\label{figcont1}
\end{figure*}

These properties are responsible for the competing large deviation scalings of $P(A,B)$ announced in the introduction. We shall show in the next sections that these scalings arise essentially from two very different fluctuation dynamics around $\ell_s$ and $\ell_u$, which will be studied analytically. To support our results, we shall also present numerical results obtained by directly integrating the Fokker-Planck equation associated with Eq.~(\ref{eqAB1}) for different noise powers.\footnote{The numerical integration was done with the routine \texttt{NDSolve} of Mathematica using double-digit accuracy and standard vanishing boundary conditions for square domains of the $A$-$B$ plane. The domains used were $[-5,5]^2$ for $\nu=0.5$ and $\nu=0.1$, $[-4,4]^2$ for $\nu=0.05$, and $[-2,2]^2$ for $\nu=0.025$. Only a portion of these regions is shown in Fig.~\ref{figcont1}.} Some of these results are shown in Fig.~\ref{figcont1} as contour plots of $P(A,B)$. We can already see from this figure that $P(A,B)$ concentrates on the stable line $\ell_s$ as $\nu\ra 0$, and that it is otherwise relatively isotropic away from $\ell_s$. Our analytical study of the $AB$ model and of its fluctuations around $\ell_s$ and $\ell_u$ will explain these two properties, among others. The main result that we shall obtain for $\sigma_A=\sigma_B=1$ is the large deviation result
\be
\lim_{\nu\ra 0} - \sqrt{\nu} \ln P(x)= I(A,B),
\ee
with
\be
I(A,B)=\frac{2\sqrt{2}}{3} (A^2+B^2)^{3/4},
\ee
away from the stable line $\ell_s$. We shall also describe corrections of this large deviation result for finite values of $\nu$ near $\ell_s$.

\section{Path integral solution}
\label{secpi}

We give in this section a first derivation of the large deviation form of $P(A,B)$ following the classical theory of Freidlin and Wentzell \cite{freidlin1984}. This calculation is valid near the line $\ell_s$ of stable points and yields the large deviation speed $c_\nu=1/\nu$.

\subsection{Classical theory}

Consider a diffusion process $\{X(t)\}$ in $\reals^D$ which is the solution of the SDE,
\be
\D X(t) =f(X(t))\, \D t+\sqrt{\nu}\, \D W(t),
\label{eqsde1}
\ee
where $\D W(t)$ are increments of the Brownian motion. It is known 
from the work of Freidlin and Wentzell (F-W) that the stationary density $P(x)$ of this process satisfies, under certain conditions, a large deviation form or \emph{large deviation principle} (LDP) in the low-noise limit $\nu\ra\infty$, which we informally write as
\be
P(x)\approx \E^{-V(x)/\nu},
\ee
to mean
\be
\lim_{\nu\ra 0} - \nu \ln P(x)= V(x).
\ee
Moreover, from the F-W theory, it is known that the \emph{rate function} or \emph{pseudo-potential} $V(x)$ defined by this limit can be obtained from the following minimization problem:
\be
V(x)=\inf_{t>0}\, \inf_{x(0)\in O, x(t)=x} L[x]
\label{eqgenwk1}
\ee
which involves the \emph{Lagrangian} or \emph{action}
\be
L[x]=\int_0^t \mL(\dot x,x)\, \D s,\qquad \mL(\dot x,x)=\frac{1}{2}\big(\dot x-f(x)\big)^2,
\ee
associated with a path $\{x(s)\}_{s=0}^t$ of the process. The minimization in (\ref{eqgenwk1}) is performed over all paths starting on the attractor $O$ of the system at time $t=0$ and reaches the point $x$ after a time $t$ which is usually taken to go to infinity.

The LDP for $P(x)$ is akin to the semi-classical or WKB approximation of quantum mechanics, and can be explained heuristically as in quantum mechanics by expressing $P(x)$ in path integral form:
\be
P(x)=\lim_{t\ra\infty} P(x,t|x\in O,0)=\lim_{t\ra\infty} \int_{x(0)\in O}^{x(t)=x} \mathcal{D}[x]\, \E^{-L[x]/\nu},
\ee
and by arguing that the probability to reach a point $x$ is given, in the low-noise limit $\nu\ra 0$, by the most probable path, called the \emph{optimal path} or \emph{instanton}, which starts on the attractor and reaches that point after a very long time. As this optimal path must have a minimal action under the terminal constraint $x(0)\in O$ and $x(t)=x$, we recover the result of (\ref{eqgenwk1}).

Freidlin and Wentzell \cite{freidlin1984} showed that this heuristic argument based on path integrals is rigorously valid for the SDE (\ref{eqsde1}) provided essentially that $O$ is a unique point attractor of the deterministic dynamics $\dot x=f(x)$. If this dynamics admits many point attractors, then the result of Eq.~(\ref{eqgenwk1}) holds locally in regions $G\subset \reals^D$ that enclose single attractors and do not include characteristic boundaries, such as separatrices. Graham \cite{graham1989,graham1995} also showed semi-heuristically that the F-W theory can be applied for non-point attractors, e.g., limit cycles or strange attractors.  

Here we apply the F-W theory to the $AB$ model in a local sense by considering a region of the upper-half plane $B>0$ surrounding the attracting line $\ell_s$ (see gray region in Fig.~\ref{figvf1}). As we take the limit $\nu\ra 0$, we neglect the dissipative terms $-\nu A$ and $-\nu B$ in the full action of the model. Therefore, the action that we consider in the minimization problem is 
\be
L[A,B]=\int_0^t\mL\,\D s,\quad \mL=\frac{1}{2\sigma_A^2} (\dot A+AB)^2+\frac{1}{2\sigma_B^2}(\dot B-A^2)^2.
\label{eqlag1}
\ee
The reason for neglecting the dissipative terms in the action is that they are sub-dominant in the low-noise limit compared with the other terms, and lead to correction terms of order $\nu$ in the rate function $I(A,B)$. The net effect of neglecting these terms when solving the minimization problem of (\ref{eqgenwk1}) is to look for instantons that emanate anywhere from the stable line $\ell_s$ of fixed points, which is the attractor of noiseless $AB$ dynamics \emph{without} the dissipation forces, i.e., Eq.~(\ref{eqdets1}), rather than emanating from the origin $(0,0)$, which is the attractor of the noiseless $AB$ dynamics \emph{with} the dissipative forces.

To find the instantons emanating from $\ell_s$, we can solve the Euler-Lagrange equations
\begin{eqnarray}
\frac{\D}{\D s}\frac{\p \mL}{\p\dot A}-\frac{\p \mL}{\p A}&=0\nonumber\\
\frac{\D}{\D s}\frac{\p \mL}{\p\dot B}-\frac{\p \mL}{\p B}&=0,
\end{eqnarray}
where $\mL$ is the Lagrangian density defined in Eq.~(\ref{eqlag1}), with the boundary conditions $A(0),B(0)\in \ell_s$ and $A(t)=A$, $B(t)=B$. By finding the solution of this equation for an increasing ``hitting'' time $t$, we then find a succession of approximations of the rate function $I(A,B)$, which converge to $I(A,B)$ as $t\ra\infty$.

For the $AB$ model, the Euler-Lagrange equations read explicitly
\begin{eqnarray}
\frac{1}{\sigma_A^2}(\ddot A+A\dot B-AB^2)-\frac{2}{\sigma_B^2}(A^3-A\dot B) &=  0 \nonumber\\
\frac{1}{\sigma_A^2}(A^2B+A\dot A)-\frac{1}{\sigma_B^2}(\ddot B-2A\dot A) &= 0.
\end{eqnarray}
This is a set of nonlinear, second-order equations of motion, which can be solved only numerically. In practice, it is easier to solve these equations by transposing them into an equivalent set of first-order Hamiltonian equations of motion defined by
\be
\dot A=\frac{\p\mH}{\p\rho_A},\quad \dot B=\frac{\p\mH}{\p\rho_B},\qquad \dot\rho_A=-\frac{\p\mH}{\p A},\quad \dot\rho_B=-\frac{\p\mH}{\p B},
\ee
where
\be
\rho_A=\frac{\p\mL}{\p\dot A},\quad \rho_B=\frac{\p\mL}{\p\dot B}
\ee
are the conjugate momenta associated with $A$ and $B$, respectively, and
\be
\mH=\rho_A \dot A+\rho_B \dot B-\mL
\ee
is the Hamiltonian density. For the $AB$ model, the Hamiltonian equations are explicitly
\begin{eqnarray}
\dot A &= \sigma_A^2\rho_A-AB\nonumber\\
\dot B &= \sigma_B^2\rho_B+A^2\nonumber\\
\dot \rho_A &= \rho_AB-2\rho_BA\nonumber\\
\dot \rho_B &= \rho_A A.
\label{eqH1}
\end{eqnarray}

\begin{figure}
\centering
\resizebox{2.2in}{!}{\includegraphics{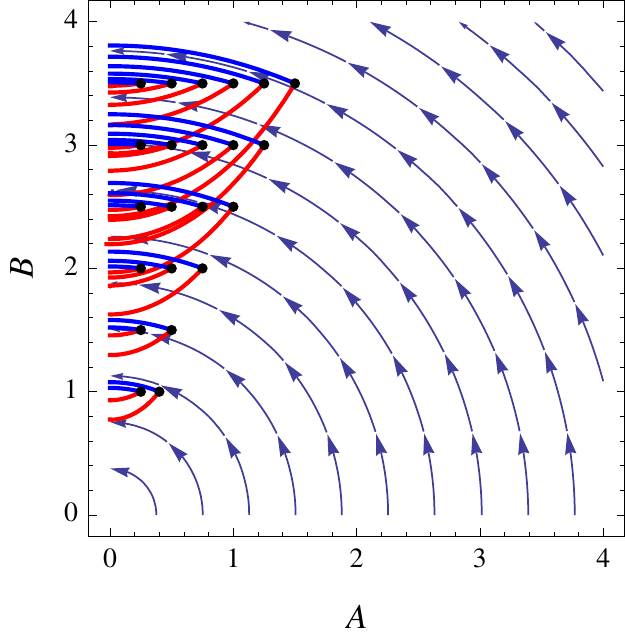}}
\caption{Red lines: Instantons reaching certain points $(A,B)$ (black dots) from the stable attracting line of the $AB$ model. Blue lines: Deterministic paths connecting the same point back to the stable attractor according to the noiseless dynamics of the $AB$ model. Parameters: $\sigma_A=\sigma_B=1$.}
\label{figcritpaths1}
\end{figure}

To numerically solve these equations, we use the fact that $\mH$ is conserved in time and that $\mH=0$ on the attractor $\ell_s$. As a result, instantons are such that $\mH=0$ for all times. Moreover, since the dynamics from the attractor is unstable, we numerically integrate the Hamiltonian equations backward in time instead of forward in time. This means that we start the integration from the point $(A,B)$, and find initial conditions for $\rho_A$ and $\rho_B$ such that the time-reverse dynamics of (\ref{eqH1}) leads to the attractor $\ell_s$.

\begin{figure}
\centering
\resizebox{3.1in}{!}{\includegraphics{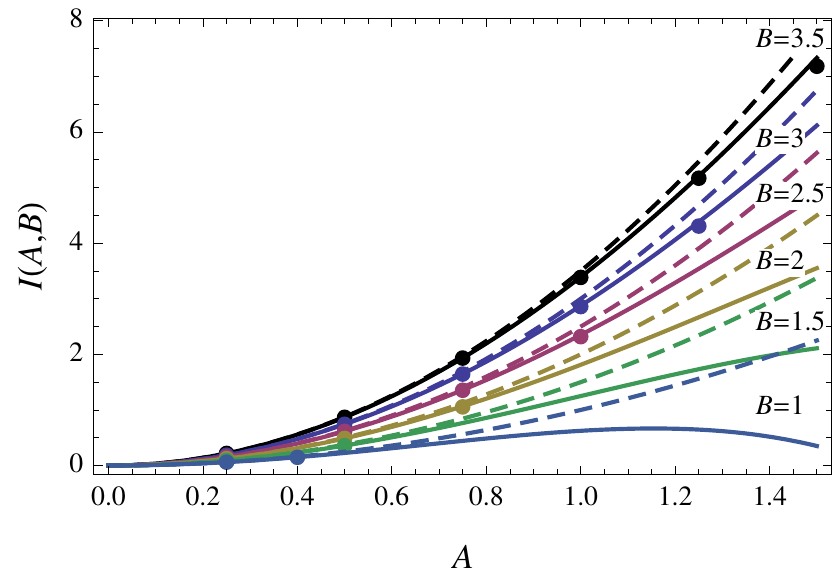}}
\caption{Rate function $I(A,B)$ near $\ell_s$ as a function of $A$ for different values of $B$. Data points: Rate function obtained from the instantons shown in Fig.~\ref{figcritpaths1}. Dashed and full lines: Second- and fourth-order approximations, respectively, of $I(A,B)$ obtained from the Hamilton-Jacobi method; see Sec.~\ref{sechj}. Parameters: $\sigma_A=\sigma_B=1$.}
\label{figres1}
\end{figure}

A number of instantons obtained from this procedure are shown in Fig.~\ref{figcritpaths1} for different points $(A,B)$ in the first quadrant of the $A$-$B$ plane.\footnote{By symmetry of the $AB$ model, instantons for the second quadrant, i.e., on the left of $\ell_s$, must be the mirror images of the instantons of the first quadrant.} The action associated with these paths, which yields the rate function $I(A,B)$ according to Eq.~(\ref{eqgenwk1}), is shown in Fig.~\ref{figres1} as data points. The speed associated with this rate function is $1/\nu$. In the same figure, we show with the full and dashed lines the results of analytical calculations presented in Sec.~\ref{sechj}. The agreement between the two sets of results will be discussed in more detail in these sections.

An interesting property of the instantons, seen from Fig.~\ref{figcritpaths1}, is that they are different from the \emph{natural decay paths} of the system, i.e., the paths of the deterministic dynamics starting from a point $(A,B)$ away from the attractor and ending on the attractor (shown in blue). This difference is characteristic of SDEs that are not \emph{gradient}, i.e., which cannot be written in the form
\be
\dot x=-\nabla U(x)+\sqrt{\nu}\,\xi(t)
\ee
for a scalar function $U(x)$. It is known that if such a form exists, then the rate function $I(x)$ associated with the stationary distribution $P(x)$ is simply given by $I(x)=2U(x)$. In this case, we moreover have that the instantons are the time-reversed version of the natural decay paths.

\begin{figure}[t]
\centering
\resizebox{2.6in}{!}{\includegraphics{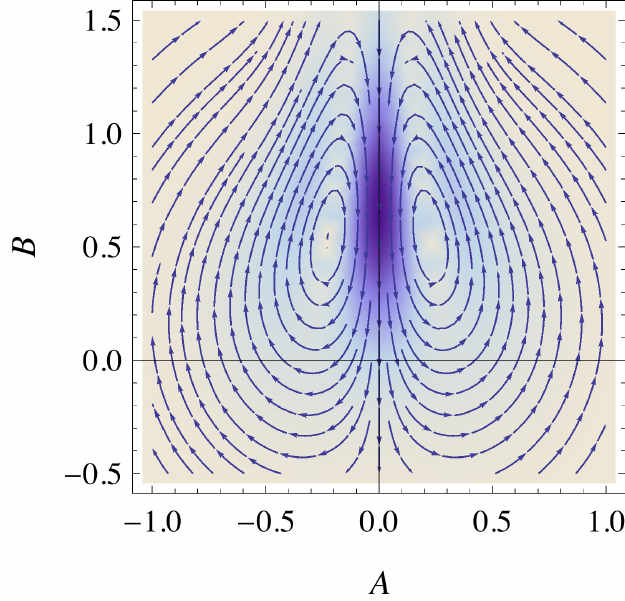}}
\caption{Streamlines of the probability current $J$ for $\nu=0.05$ and $\sigma_A=\sigma_B=1$, with level colors representing $|J|$. The darker colors correspond to larger current magnitudes.}
\label{figcurr1}
\end{figure}

For the $AB$ model, these results do not apply: the SDE (\ref{eqAB1}) is not of gradient-type, which means that we must explicitly solve the Euler-Lagrange equations or the equivalent Hamiltonian equations to find the instantons and their associated action. Physically, this also means that the $AB$ model is a nonequilibrium system that violates detailed balance and gives rise to a non-vanishing probability current $J=(J_A,J_B)^T$, whose components are here given by
\begin{eqnarray}
J_A &=(-AB-\nu A)P(A,B)-\frac{\nu\sigma_A^2}{2}\frac{\p P(A,B)}{\p A} \nonumber\\
J_B &=(A^2-\nu B)P(A,B)-\frac{\nu\sigma_B^2}{2}\frac{\p P(A,B)}{\p B}.
\label{eqcurr1}
\end{eqnarray}
This current is actually a current loop, as shown in Fig.~\ref{figcurr1}: it has the shape of a ``squashed'' double-solenoid and circulates through the origin. 

\subsection{Sub-instantons}

\begin{figure}
\centering
\resizebox{2.5in}{!}{\includegraphics{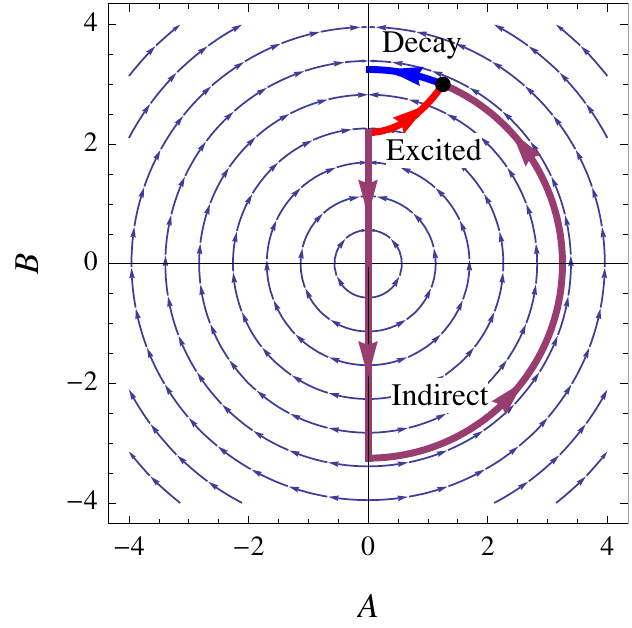}}
\caption{Optimal paths. The point in black in the $A$-$B$ plane can be reached either by a direct, excited path (in red) with positive action or by an indirect path (in purple) that first goes along the stable and unstable lines of fixed points, and then follows a natural trajectory of the deterministic dynamics. The latter path has a null action.}
\label{figcritp1}
\end{figure}

The instantons shown in Fig.~\ref{figcritpaths1} are not strictly speaking the only possible instantons that the $AB$ model admits. Interestingly, it is also possible to reach any points $(A,B)$ from the attractor by following a fluctuation path constructed as follows (see Fig.~\ref{figcritp1}):
\begin{enumerate}
\item Start from any point on the stable attractor and go down to the origin;
\item Follow the line $\ell_u$ of unstable fixed points down to the point $(0,B')$ such that $B'^2=A^2+B^2$;
\item Follow the natural orbit of energy $E=A^2+B^2$ connecting $(0,B')$ to $(A,B)$.
\end{enumerate}
As each of these parts lies either on the line of stable and unstable points or follows a natural path of the deterministic system, the complete path thus constructed must have a zero action on the large deviation scale $1/\nu$.

The existence of these zero-action instantons, or \emph{sub-instantons} as we shall call them, gives us an indication that the \emph{dominant} behavior of $P(A,B)$ does not have a speed $1/\nu$; in fact, it has a speed $1/\sqrt{\nu}$, as we shall see below. The crucial point to note, however, is that $P(A,B)$ has a sub-dominant LDP with speed $1/\nu$ near the attractor $\ell_s$ (see gray region in Fig.~\ref{figvf1}), which is precisely the LDP that we have obtained before with the F-W calculation. This sub-dominant LDP arises from instantons having a non-zero action emanating from $\ell_s$, whereas the dominant LDP, which scales with $1/\sqrt{\nu}$, is related to fluctuation paths that are sub-instantons. The fact that these sub-instantons follow in the end the deterministic dynamics explains the observed isotropy of $P(A,B)$ away from $\ell_s$, i.e., why $P(A,B)$ has constant level curves along the vector field of the $AB$ noiseless dynamics. The same indirect fluctuation paths also explain the existence of the probability current mentioned before.

We show in the next two sections how to go beyond the F-W calculation to obtain the dominant LDP of $P(A,B)$. Two methods are presented: the first is based on the solution of a time-scale separation of the solution of the AB model, whereas the second is based on a direct approximation of the solution of the Fokker-Planck equation.

\section{Dynamical analysis}
\label{secda}

We analytically derive in this section LDPs for $P(A,B)$ near $\ell_s$ and then near $\ell_u$ using the insight gained in the previous section. In treating $\ell_s$ and $\ell_u$ separately, we shall see how the stable and unstable dynamics, respectively, of each of these lines gives rise to different LDPs. As a prelude to these calculations, we obtain the stationary probability density of the energy defined by $E=A^{2}+B^{2}$ to show that it does not scale with $\nu$, as is the case for weakly-perturbed Hamiltonian systems.

\subsection{Energy}

Given the original system of equations for $A$ and $B$, displayed in Eq.~(\ref{eqAB1}), $E$ is found to evolve according to the It\^o SDE: 
\be
\D E=-2\nu E\D t+\nu\left(\sigma_{A}^{2}+\sigma_{B}^{2}\right)\D t+2\sqrt{\nu}\left(\sigma_{A}A\,\D W_{A}+\sigma_{B}B\,\D W_{B}\right).
\label{eq:Energy_Evolution}
\ee
This equation is not closed -- it explicitly depends on $A$ and $B$ -- but the equation for the mean energy is closed:
\be
\frac{1}{\nu}\frac{d\left\langle E\right\rangle }{dt}=-2\left\langle E\right\rangle +\left(\sigma_{A}^{2}+\sigma_{B}^{2}\right).
\ee
From this, we find the stationary mean energy:
\be
\left\langle E\right\rangle _{s}=\frac{\sigma_{A}^{2}+\sigma_{B}^{2}}{2}.
\label{eq:Stationary_Average_Energy}
\ee

To find $P(E)$, we solve the Fokker-Planck equation associated with the Langevin equation for $E$. For simplicity, consider first the case $\sigma_A=\sigma_B=\sigma$ and let $\phi$ be any test function. Applying It\^o's formula and averaging over the noises, we obtain 
\begin{equation}
\frac{1}{2\nu}\frac{\D\left\langle \phi(E)\right\rangle }{\D t}=\left\langle \left(\sigma^{2}-E\right)\phi'(E)\right\rangle
+\sigma^{2}\left\langle E\phi''(E)\right\rangle.
\label{eq:test_function_E}
\end{equation}
The associated Fokker-Planck equation for the energy pdf $P(E,t)$ is thus
\be
\frac{1}{2\nu}\frac{\p P}{\p t}=
\frac{\p}{\p E}\left[\left(E-\sigma^{2}\right)P+\sigma^{2}\frac{\p}{\p E}(EP)\right],
\ee
and admits the following stationary solution $P(E)$:
\be
P(E)=\frac{1}{\sigma^{2}}\E^{-E/\sigma^2}.
\label{eqfps1}
\ee
It is easy to verify from this result that $\left\langle E\right\rangle_{s}=\sigma^{2}$, in accordance with the exact result of Eq.~(\ref{eq:Stationary_Average_Energy}). Note also that $P(E)$ does not depend on $\nu$, as announced, which means that there is no LDP for $P(E)$.

For the case $\sigma_A\neq\sigma_B$, It\^o's formula leads to 
\be
\frac{1}{2\nu}\frac{\D\left\langle \phi\left(E\right)\right\rangle }{\D t}=\left\langle \left(\frac{\sigma_{A}^{2}+\sigma_{B}^{2}}{2}-E\right)\phi'(E)\right\rangle
+\left\langle \left(\sigma_{A}^{2}A^{2}+\sigma_{B}^{2}B^{2}\right)\phi''(E)\right\rangle,
\label{eq:test_function_E_2}
\ee
which is exact for any test function $\phi$. By contrast with Eq.~(\ref{eq:test_function_E}), the equation above is not a closed equation for $E$ as the variables $A$ and $B$ are involved. However, because the typical values of the variable $A$ are of order $\sqrt{\nu}\sigma_{1}/\sqrt{B}$, which is much smaller than both $\sqrt{E}$ and $B$ as soon as $E\gg\nu^{2}\sigma_{1}^{4}$, it is natural in Eq.~(\ref{eq:test_function_E_2}) to make the approximations $B^{2}\simeq E$ and $A^{2}\ll E$. One then obtains a closed equation for $\phi(E)$ whose associated Fokker-Planck equation is
\be
\frac{1}{2\nu}\frac{\p P}{\p t}=\frac{\p}{\p E}\left[\left(E-\frac{\sigma_{A}^{2}+
\sigma_{B}^{2}}{2}\right)P+\sigma_{B}^{2}\frac{\p}{\p E}(EP)\right].
\label{eq:Fooker_Planck_E_aproximation}
\ee
The stationary solution of this equation is
\be
P(E)=C\, E^{(\sigma_{A}^{2}-\sigma_{B}^{2})/(2\sigma_{B}^{2})}\, \E^{-E/\sigma_{B}^{2}},
\label{eq:PDF_Energy_Approximation}
\ee
where $C$ is a normalization constant. We can verify that this expression reduces to the one we found in Eq.~(\ref{eqfps1}) when $\sigma_{A}=\sigma_{B}=\sigma$. It is also easily verified that the average energy obtained from this distribution is given by the exact relation displayed in Eq.~(\ref{eq:Stationary_Average_Energy}). 

\subsection{Dynamics around stable fixed points}

We now derive a local LDP for $P(A,B)$ around $\ell_s$ (see gray region in Fig.~\ref{figvf1}) using the fact that there exists a separation between the time with which $A$ relaxes to $0$ (time of order $1/B$) and the time with which $E$ varies (time of order $1/\nu$). Since $E\approx B^2$ for $A$ close to $0$, the slow dynamics of $E$ translates into a slow dynamics of $B$, which can be treated adiabatically with respect to $A$. Thus, we consider $B$ to be constant and study the dynamics of the rapid variable $A$ given by the first equation of the $AB$ model:
\be
\D A =  (-AB-\nu A)\,\D t+\sigma_{A}\sqrt{\nu}\,\D W_{A}
\ee
For fixed $B$, this equation is linear in $A$, which means that $A$ evolves according to a Ornstein-Uhlenbeck process. In the stationary state, $A$ is of order $\sqrt{\nu}$ and the term $\nu A$ can be considered negligible, and so we find
\be
P(A|B)=\sqrt{\frac{B}{\pi\nu \sigma_A^2}}\,\exp{\left(-\frac{BA^2}{\nu\sigma_A^2}\right)},
\ee
for the stationary distribution of the fast variable $A$ given that $B$ is fixed. This stationary distribution is reached on a time-scale of order $1/B$.

Next, we study the evolution of $B$ given by
\be
\D B=\nu (A'^2-B)\D t+\sqrt{\nu}\, \sigma_B\, \D W_B,
\ee
using $A'=A/\sqrt{\nu}$. As we did for $A$, we can approximate $A'$ to be an Ornstein-Uhlenbeck process with variance $\langle A'^2\rangle=\sigma_A^2/(2B)$. Accordingly, the fluctuations of the variable $\nu A'^2$ must be such that $\langle (\nu A'^2-\langle \nu A'^2\rangle)^2\rangle$ is of order $\nu^2$. Such fluctuations are much smaller than the effect of the white noise for $B$, and so, to leading order in $\nu$, the fluctuations of $A'^2$ can be neglected. The approximate dynamics for $B$ to leading order in $\nu$ is thus
\be
\D B=\nu\left(\frac{\sigma_A^2}{2B}-B \right)\D t+\sqrt{\nu}\,\sigma_B\, \D W_B.
\ee
The associated Fokker-Planck equation is 
\be
\frac{1}{\nu}\frac{\p P}{\p t}=\frac{\p}{\p B}\left[ \left(B-\frac{\sigma_A^2}{2B}\right)P+\frac{\sigma_B^2}{2}\frac{\p P}{\p B}\right],
\ee
and has for stationary solution
\be
P(B)=C\, B^{\sigma_A^2/\sigma_B^2}\, \E^{-B^2/\sigma_B^2},
\ee
where $C$ is a normalization constant. By combining this solution with the solution found for $P(A|B)$, we thus find
\be
P(A,B) =P(A|B)P(B)
=C\, B^{\frac{2\sigma_A^2+1}{2\sigma_B^2}} \exp\left(-\frac{B^2}{\sigma_B^2}-\frac{BA^2}{\nu \sigma_A^2}\right),
\label{eqdynstable1}
\ee
where $C$ is again a normalization constant. The LDP extracted from this result has the form
\be
P(A,B)\approx \E^{-I(A,B)/\nu},\qquad I(A,B)=-\frac{BA^2}{\sigma_A^2},
\label{eqldps1}
\ee
and is nothing but the LDP associated with $P(A|B)$. This shows that the large deviations of the $AB$ model around the attractor are mainly the result of the diffusive fluctuations of $A$ for $B$ constant.

The rate function $I(A,B)$ of Eq.~(\ref{eqldps1}) is shown in Fig.~\ref{figres1} as a function of $A$ and for different values of $B$ as dashed lines. We see that there is a favorable comparison between this rate function and the rate function obtained from the F-W theory (represented again by the data-points). The difference with the F-W results can be explained by noting that the rate function derived in this section is only a second-order approximation of the rate function obtained by approximating the instantons as straight, horizontal paths joining a point $(0,B)$ on $\ell_s$ to a point $(A,B)$ near $\ell_s$. The correct instantons, as seen from Fig.~\ref{figcritpaths1}, have a certain inclination or curvature in the $B$ direction, which has the effect of reducing slightly the action of straight, constant-$B$ paths. We shall see in Sec.~\ref{sechj} that the Hamiltonian-Jacobi method is able to capture this effect as a  fourth-order correction to $I(A,B)$.

\subsection{Dynamics around unstable fixed points}

To obtain $P(A,B)$ around the line $\ell_u$ of unstable fixed points, we need to construct a different approximation of the $AB$ dynamics that accounts for the fact that trajectories starting at the origin have a small probability of diffusing down the line $\ell_u$ and that, as soon as they depart from this line, they are rapidly expelled along the force lines of the $AB$ dynamics, to reach the stable line $\ell_s$; see Figs. \ref{figvf1} and \ref{figcritp1}. This dynamics creates a probability current because, eventually, the trajectories return to the origin by diffusing down the line $\ell_s$ and start new loops by diffusing again down $\ell_u$. 

In order to qualitatively understand this fluctuation dynamics, we consider the following approximation of the $AB$ model:
\begin{eqnarray}
\D A & =  \alpha_A A \D t+\sigma_{A}\sqrt{\nu}\,\D W_{A}\nonumber\\
\D B & =  -\nu B\,\D t+\sigma_{B}\sqrt{\nu}\,\D W_{B},
\label{eqABu1}
\end{eqnarray}
which involves a ``top-hill'' diffusion process for $A$ ($\alpha_A>0$) and a ``down-hill'' diffusion for $B$. The equation for $B$ is the exact dynamics of $B$ on the line $A=0$, whereas the ``top-hill'' term $\alpha_A A$ is analogous to the term $-AB$ in the original $AB$ model.  For now, we consider $\alpha_A$ to be independent of $B$ in order to have a solvable model which will help us gain a qualitative understanding of the unstable dynamics. We discuss the case $\alpha_A=-B$ at the end of this section.

For  $\alpha_A$ constant, Eq.~(\ref{eqABu1}) describes an unstable linear Gaussian process. To find a stationary distribution $P(A,B)$ for this process, we have to consider that there is a probability (or current) source at the origin $O$. In this case, the probability to reach a point $(A,B)$ in the lower plane is given by integrating the probability $P(A,B,t|O)$ that a trajectory starting at the origin $O$ reaches the point $(A,B)$ after a variable time $t$, to which we multiply the stationary probability density of starting at $O$:
\be
P(A,B)= P(O) \int_0^\infty P(A,B,t|O)\,\D t.
\ee
For the decoupled dynamics of Eq.~(\ref{eqABu1}), the propagator $P(A,B,t|O)$ is simply given by
\be
P(A,B,t|O)=\frac{1}{2\pi \sqrt{\langle A^2\rangle\langle B^2\rangle}} \exp\left[-\left(\frac{A^2}{2\langle A^2\rangle}+\frac{B^2}{2\langle B^2\rangle}\right)\right],
\ee
where
\be
\langle A^2\rangle=\frac{\sigma_A^2\nu}{2\alpha_A}(\E^{2\alpha_A t}-1)
\ee
and
\be
\langle B^2\rangle=\frac{\sigma_B^2}{2}(1-\E^{-2\nu\sigma_B^2 t}).
\ee

The difference between the behavior of the variance of $A$ and $B$ plays an important role in determing the LDP of $P(A,B)$ in the low-noise limit. The variance of $A$ grows exponentially for large times, as a result of the instability of $\ell_u$, whereas that of $B$ goes to a constant on a long time-scale of order $t\sim (2\sigma_B^2 \nu)^{-1}$. Taking the limit $\nu\ra 0$, and anticipating that the integral will be dominated by contributions with times such that $\nu t \ll 1$, we obtain
\be
P(A=0,B)\approx C\int_0^\infty \E^{-\alpha_A t-B^2/(2\sigma_B^2\nu t)}\, \D t,
\ee
for the probability density on $\ell_u$, with $C$ a normalization constant. This can be put in the form of a Laplace integral with the change of variables $t'=\sqrt{\sigma_B^2\nu}\, t$:
\be
P(A=0,B)\approx C\int_0^\infty \E^{-g(t')/\sqrt{\nu}}\, \D t'
\ee
where
\be
g(t')=\frac{\alpha_A}{\sigma_B} t'+\frac{B^2}{2\sigma_B t'}.
\ee
In the limit $\nu\ra 0$, the integral is therefore dominated by its maximum integrand, which is here located at $t'=B/\sqrt{2\alpha_A}$. As a result, we find
\be
\label{phi}
P(A=0,B)\approx \E^{-\varphi(B)/\sqrt{\nu}}, \qquad \varphi(B)=\frac{\sqrt{2\alpha_A}}{\sigma_B} B.
\ee
Note the change of speed in this LDP: $1/\sqrt{\nu}$ instead of $1/\nu$.

From this simple ``top-hill'' linear model, we conclude that the low probability $P(0,B)$ for the system to be on the unstable line $\ell_u$ results from the balance between the probability to reach a finite value $B$ after a time $t$ and the probability that the system is not expelled by the unstable dynamics before this time $t$. The rare coincidence of these two conditions gives the large deviation result. The preceding computation suggests that the leading contribution corresponding to those rare events correspond to an optimal time $t(B)$ proportional to $\sqrt{\nu}$. 

This qualitative behavior also correctly describes the exact $AB$ dynamics with $\alpha_A = B$. In this case, one can explicitly solve the two equations shown in~(\ref{eqABu1}) with $\alpha_A = B$. However, we have found more convenient to use in this case an asymptotic expansion of the Hamilton-Jacobi equation, which we discuss in the next section. For now, we just note that substituting $\alpha_A=B$ in Eq.~(\ref{phi}) leads to $\varphi(B) \propto B^{3/2}/\sigma_B$, a result that will be confirmed in the next section.

To close this section, note that the density $P(A,B)$ away from $\ell_u$ is obtained by radially radiating the value of $P(0,B)$ obtained above, i.e., by setting $P(A',B')=P(0,B)$ for all $(A',B')$ such that $A'^2+B'^2=B^2$. This is expected from the numerical results shown in Fig.~\ref{figcont1} and follows, more precisely, by noticing that trajectories departing infinitesimally from $\ell_u$ are radiated radially by the vector field of the $AB$ model. As this instability results only from the deterministic part of the $AB$ equations, the probability density must have circular level curves.

\section{Hamilton-Jacobi approach}
\label{sechj}

We give in this section a different derivation of the LDP of $P(A,B)$ near $\ell_s$ and $\ell_u$, obtained by solving the stationary Fokker-Planck equation of the $AB$ model, which reads
\be
\fl
\frac{\p}{\p A} (AB P)-\frac{\p}{\p B}(A^2P)+\nu\frac{\p}{\p A}(AP)+\nu\frac{\p}{\p B}(BP)
+\nu\frac{\sigma^2_A}{2}\frac{\p^2 P}{\p A^2}+\nu\frac{\sigma^2_B}{2}\frac{\p^2 P}{\p B^2}=0.
\label{eqfp1}
\ee
As before, we look for solutions having the LDP form
\be
P(A,B)=\phi(A,B)\,\E^{-c_\nu I(A,B)},
\label{eqans1}
\ee
where $I(A,B)$ is the rate function and $\phi(A,B)$ a function of $A$ and $B$ that does not depend on $\nu$. Moreover, following the previous section, we choose the noise scaling $c_\nu=1/\nu$ around $\ell_s$ and $c_\nu=1/\sqrt{\nu}$ around $\ell_u$. 

The advantage of the Fokker-Planck method is that it yields more precise expressions for the rate function in the form of series expansions, which are not based, as in the previous section, on approximations of the $AB$ model. However, to be able to use this method, we need to know the correct large deviation speed of $P(A,B)$, i.e., the correct scaling of $c_\nu$ with $\nu$.

\subsection{Stable fixed points}

Inserting the ansatz (\ref{eqans1}) with $c_\nu=1/\nu$ in the Fokker-Planck equation (\ref{eqfp1}) yields, to order $1/\nu$, the following partial differential equation:
\be
-AB\frac{\p I}{\p A}+A^2\frac{\p I}{\p B}+\frac{\sigma_A^2}{2}\left(\frac{\p I}{\p A}\right)^2+\frac{\sigma_B^2}{2}\left(\frac{\p I}{\p B}\right)^2=0.
\label{eqhj1}
\ee
This is a Hamilton-Jacobi equation corresponding to the weak-noise limit of our model. Unfortunately, we are unable to solve this equation explicitly. However, we expect the solution to be localized close to $A=0$, and so it is natural to look for an expansion of $I$ for small $A$. By symmetry of the $AB$ model, the dependence of $P(A,B)$ on $A$ must be even. Thus, assuming an expansion of $I$ of the form
\be
I(A,B)=I_{0}(B)+I_{1}(B)A^2+I_{2}(B)A^4+O(A^6),
\ee
in Eq.~(\ref{eqhj1}), we directly obtain
\be
I(A,B)=\frac{B}{\sigma_A^2}A^2-\frac{2\sigma_A^2+\sigma_B^2}{8\sigma_A^4B}A^4+O(A^6).
\label{eqphi01}
\ee

Figure~\ref{figres1} shows how this result compares with the calculation of the rate function based on the instanton approximation presented in Sec.~\ref{secpi}. The full colored lines on this plot show $I(A,B)$, as given in Eq.~(\ref{eqphi01}), as a function of $A$ for different values of $B$. The dashed colored lines represent the same result but truncated to second order. The data points, as mentioned in Sec.~\ref{secpi}, are the results of the instanton approximation. The near perfect match between the data points and the full lines confirm the consistency of the Hamilton-Jacobi and instanton results. 

\begin{figure}
\centering
\resizebox{3.1in}{!}{\includegraphics{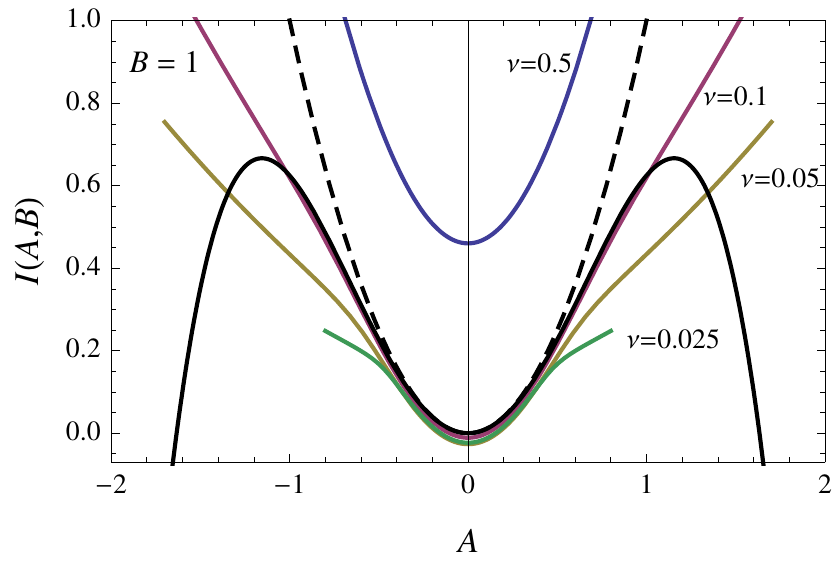}}
\caption{Numerical test for $I(A,B)$. Black dashed line: Second-order result for $I(A,B)$ obtained from the Hamilton-Jacobi method. Black full line: Fourth-order analytical result for $I(A,B)$ obtained with the same method. Colored lines: Rate functions obtained from the numerical integration of the Fokker-Planck equation of the $AB$ model for different values of $\nu$. Parameters: $\sigma_A=\sigma_B=1$.}
\label{figres2}
\end{figure}

The result of Eq.~(\ref{eqphi01}) is also compared in Fig.~\ref{figres2} with results obtained by numerically integrating the Fokker-Planck equation. We see in this figure that, for small values of $\nu$, the numerical results are relatively close to $I$, at least for points $(A,B)$ near $\ell_s$. It must be said that it is rather difficult to obtain reliable results by numerically integrating the Fokker-Planck with small noise intensities, especially away from the attractor. In our case, we could not obtain reliable results for values of $\nu$ smaller than about $\nu=0.025$. For this reason, the numerical results of Fig.~\ref{figres2} are presented to check that our analytical results have the correct scaling in $\nu$, and that the rate functions obtained numerically and analytically are qualitatively similar.

To obtain the correction factor $\phi$, we need to consider the Hamilton-Jacobi equation for $\phi$ obtained at order $0$ of the Fokker-Planck equation:
\begin{eqnarray}
\fl
B\phi+AB\frac{\p\phi}{\p A}-A^2\frac{\p\phi}{\p B}-A\frac{\p I}{\p A}\phi-B\frac{\p I}{\p B}\phi\nonumber\\
-\frac{\sigma_A^2}{2}\left(\frac{\p^2 I}{\p A^2}\phi+2\frac{\p I}{\p A}\frac{\p\phi}{\p A}\right)
-\frac{\sigma_B^2}{2}\left(\frac{\p^2 I}{\p B^2}\phi+2\frac{\p I}{\p B}\frac{\p\phi}{\p B}\right)=0.
\label{eqhj2}
\end{eqnarray}
As before, we attempt to solve this equation by expanding $\phi$ in even powers of $A$. Thus we assume
\be
\phi(A,B)=\phi_{0}(B)+\phi_{1}(B)A^2+O(A^4),
\label{eqtayl1}
\ee
and find at order $A^2$, 
\be
4(\sigma_A^2+\sigma_B^2)B\phi_{0}'+(12B^2-6\sigma_A^2-3\sigma_B^2)\phi_{0}+8B^2\sigma_A^2\phi_{1}=0,
\label{eqhj3}
\ee
where $\phi_{0}'(B)$ is the $B$-derivative of $\phi_{0}(B)$. This equation is not a closed differential equation for $\phi_{0}$ as it involves $\phi_{1}$. To get a closed set of equations, we find a complementary equation by looking at the Hamilton-Jacobi equation obtained at order $\nu$, which reads
\begin{eqnarray}
\fl
B\varphi+AB\frac{\p\varphi}{\p A}-A^2\frac{\p\varphi}{\p B}+2\phi-A\frac{\p I}{\p A}\varphi
+A\frac{\p\phi}{\p A}-B\frac{\p I}{\p B}\varphi+B\frac{\p\phi}{\p B}\nonumber\\
-\frac{\sigma_A^2}{2}\left(\frac{\p^2 I}{\p A^2}\varphi+2\frac{\p I}{\p A}\frac{\p\varphi}{\p A}-\frac{\p^2\phi}{\p A^2}\right)\nonumber\\
-\frac{\sigma_B^2}{2}\left(\frac{\p^2 I}{\p B^2}\varphi+2\frac{\p I}{\p B}\frac{\p\varphi}{\p B}-\frac{\p^2\phi}{\p B^2}\right)=0.
\end{eqnarray}
We seek a solution of the form $\varphi(A,B)=\varphi_{0}(B)+O(A^2)$ for the function $\varphi$ carrying the $\nu$-order correction. Substituting in the equation above the expressions of $I$ and $\phi$ shown in (\ref{eqphi01}) and (\ref{eqtayl1}), respectively, we obtain at order $A^0$:
\be
\sigma_B^2\phi_{0}''+2B\phi_{0}'+4\phi_{0}+2\sigma_A^2\phi_{1}=0.
\ee
With this equation and Eq.~(\ref{eqhj3}), it is possible to eliminate the function $\phi_{1}$ to obtain a closed differential equation for $\phi_{0}$:
\be
4B^2\sigma_B^2\phi_{0}''+[8B^3-4B(\sigma_A^2+\sigma_B^2)]\phi_{0}'
+(4B^2-6\sigma_A^2-3\sigma_B^2)\phi_{0}=0.
\label{eqhj4}
\ee
Looking for solutions of the form $CB^\alpha\E^{-\beta B^2}$, we then find
\be
\phi_{0}(B)=C B^{\frac{2\sigma_A^2+1}{2\sigma_B^2}}\E^{-B^2/\sigma_B^2},
\ee
where $C$ is a normalization constant. This is the only normalizable solution to Eq.~(\ref{eqhj4}).

With the solution of $ I$ truncated to order $O(A^2)$ and $\phi$ truncated to order $O(A^0)$, i.e., $\phi=\phi_{0}+O(A^2)$, we can finally write down our approximation for $P(A,B)$:
\begin{eqnarray}
\fl
P(A,B) \approx  \left[B^{\frac{2\sigma_A^2+1}{2\sigma_B^2}}\exp\left(-\frac{B^2}{\sigma_B^2}\right)+O(A^2)\right]\nonumber\\
 \times\exp\left(-\frac{BA^2}{\nu\sigma_A^2}+\frac{2\sigma_A^2+\sigma_B^2}{8\sigma_A^4B}A^4+O(A^6/\nu)\right),
\end{eqnarray}
which reproduces the result of Eq.~(\ref{eqdynstable1}) to second order in $A$ in the exponential.

\subsection{Unstable fixed points}

\begin{figure}[t]
\centering
\resizebox{3.1in}{!}{\includegraphics{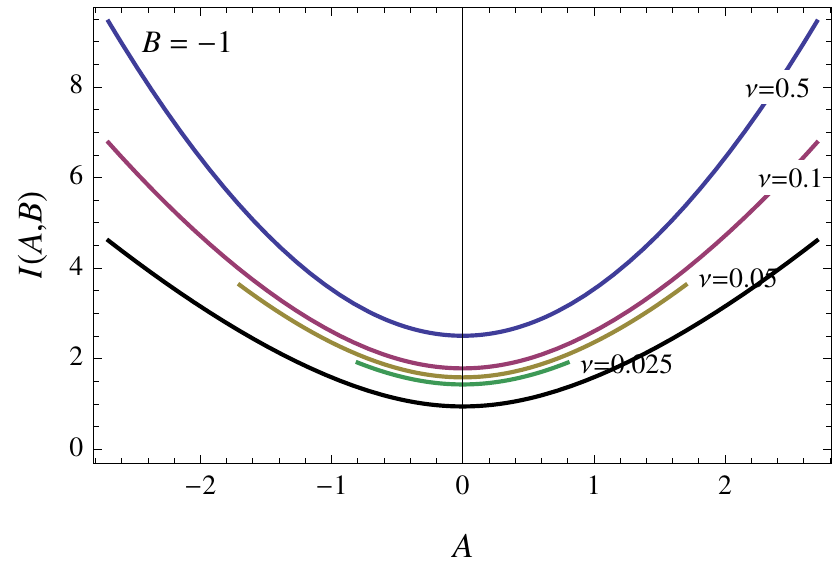}}
\caption{Numerical test for $ I(A,B)$ on the line $B=-1$. $\sigma_A=\sigma_B=1$. Black line: Analytical result for $ I(A,B)$. Colored lines: Rate function $ I(A,B)$ obtained from the numerical integration of the Fokker-Planck equation of the $AB$ model for different values of $\nu$.}
\label{figres3}
\end{figure}

Knowing from the previous sections that $P(A,B)$ is isotropic around the line $\ell_u$ of unstable points, it makes sense to solve the Fokker-Planck equation in polar coordinates:
\begin{eqnarray}
\fl
\frac{\p P}{\p t} =
-\frac{\p}{\p r}\left[ \left(-\nu r+\frac{\nu}{2r}(\sigma_A^2\sin^2\theta+\sigma_B^2\cos^2\theta)\right)P\right]\nonumber\\
 -\frac{\p}{\p\theta}\left[\left(r\cos\theta-\nu\sin 2\theta+\frac{\nu\sin 2\theta}{2r^2}(\sigma_A^2+\sigma_B^2)\right)P\right]\nonumber\\
+\frac{\nu}{2}\frac{\p^2}{\p r^2}\left[(\sigma_A^2\cos^2\theta+\sigma_B^2\sin^2\theta)P\right]\nonumber\\
 +\frac{\nu}{2}\frac{\p^2}{\p r\p\theta}\left[\left(-\sigma_A^2\frac{\sin 2\theta}{r}+\sigma_B^2\frac{\sin 2\theta}{r}\right)P\right]\nonumber\\
+\frac{\nu}{2}\frac{\p^2}{\p\theta^2}\left[\left(\sigma_A^2\frac{\sin^2\theta}{r^2}+\sigma_B^2\frac{\cos^2\theta}{r^2}\right)P\right],
\end{eqnarray}
using the large deviation ansatz $P(A,B)=\phi\E^{- I/\sqrt{\nu}}$. At order $\nu^{-1/2}$, and setting $\sigma_A=\sigma_B=1$ for simplicity, this equation reduces to the simple partial differential equation
\be
-r\cos\theta\,\frac{\p I}{\p\theta}=0,
\ee
which shows that $ I$ does not depend on $\theta$ for $A\neq 0$ (i.e., $\theta\neq\pm\pi/2$). To find the $r$-dependence of $ I$, we have to consider the Hamilton-Jacobi equation associated with the $0$th order of the Fokker-Planck equation in $\nu$, which reads
\be
-\frac{\p}{\p\theta}\left(r\,\phi\cos\theta \right)+\frac{1}{2}\left(\frac{\D I}{\D r}\right)^2\phi=0.
\ee
We look for a solution of this equation close to $\theta=-\pi/2$ by making the change of variables $\theta=-\pi/2+u$ and by defining $\tilde\phi(u,r)=\phi(-\pi/2+u,r)$. The equation above then becomes
\be
\frac{\p}{\p u}\left(r\,\tilde\phi\sin u \right)-\frac{1}{2}\left(\frac{\D I}{\D r}\right)^2\tilde\phi=0.
\ee
By symmetry, $\tilde\phi$ must be even in $u$, which implies that $\p\tilde\phi/\p u=0$ at $u=0$, and so we find
\be
\left[r-\frac{1}{2}\left(\frac{\D  I}{\D r}\right)^2\right]\tilde\phi(0,r)=0.
\ee
Assuming that $\tilde\phi(0,r)>0$, we then obtain 
\be
 I(r)=\frac{2\sqrt{2}}{3}r^{3/2}
\ee
as the solution for $ I$.\footnote{Note that $I$ must be positive and must vanish on the stable fixed point $(0,0)$.} In terms of $A$ and $B$, this yields the LDP
\be
P(A,B)\approx \E^{- I(A,B)/\sqrt{\nu}},
\ee
where
\be
I(A,B)=\frac{2\sqrt{2}}{3} (A^2+B^2)^{3/4}.
\ee

This rate function is different from the one obtained in Sec.~\ref{secda}. This is because the dynamical model studied in Sec.~\ref{secda} is an approximation of the $AB$ model in which the dynamics of $A$ is decoupled from $B$. This approximation is not assumed here, so we should take the solution above as a more precise expression of the rate function. The essential point to notice is that both approaches confirm the speed $1/\sqrt{\nu}$ of the LDP of $P(A,B)$ near $\ell_u$, which also appears to qualitatively match the scaling of the numerical Fokker-Planck results; see Fig.~\ref{figres3}. As before, it is difficult to push the numerical scaling analysis beyond $\nu=0.025$, as the numerical integration of the Fokker-Planck equation becomes unstable for smaller values of $\nu$.

\section{Discussion}
\label{secconc}

The two LDPs that we have derived for $P(A,B)$ can be thought of as an expansion of $P(A,B)$ that keeps its two first dominant terms, so that
\be
P(A,B)= C_1\, \E^{- I(A,B)/\sqrt{\nu}}+C_2\, \E^{-J(A,B)/\nu}.
\ee
As discussed, the dominant term in this expansion is the $1/\sqrt{\nu}$ term associated with the repelling line and is ``non-classical'' in the sense that it does not arise from instanton fluctuation paths that are characteristic of the theory of Freidlin and Wentzell \cite{freidlin1984}. The second term with large deviation speed $1/\nu$ is the term that arises from this theory: it is associated with instantons emanating from the attracting line and becomes important, for small but finite $\nu$, in the vicinity of this line. 

\begin{figure}
\centering
\resizebox{4in}{!}{\includegraphics{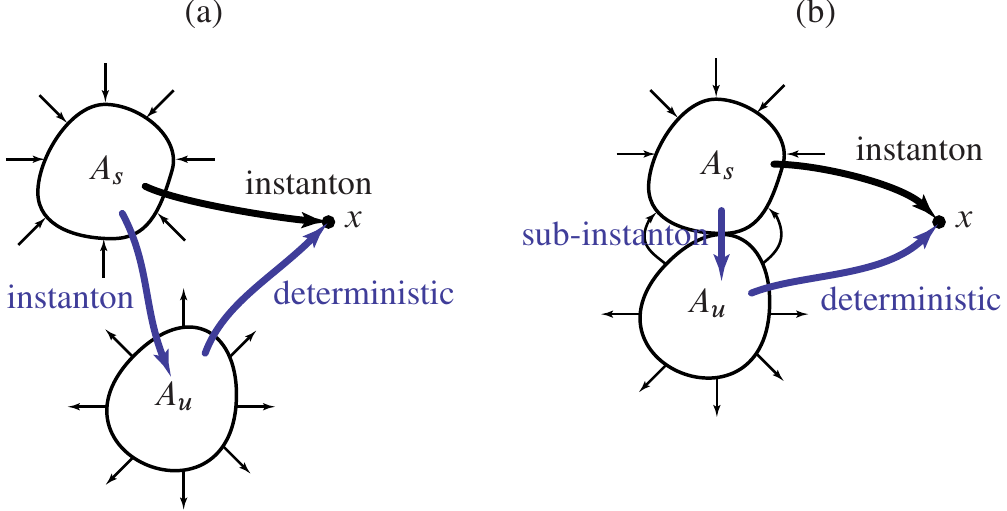}}
\caption{(a) Fluctuation paths for disconnected sets of stable ($A_s$) and unstable ($A_u$) fixed points. In this case, any path reaching a point $x$ must contain an instanton part, even if it goes through $A_u$ (blue path), so its large deviation probability must have a speed $1/\nu$. (b) Fluctuation paths for connected sets of stable and unstable fixed points. In this case, there are fluctuation paths that reach $x$ via ``sub-instantons'' (blue path) instead of instantons (black path). Sub-instantons have a large deviation speed smaller than $1/\nu$.}
\label{figgenarg1}
\end{figure}

The presence of the stable and unstable lines of fixed points is important for obtaining these results, but what is also important is that these two sets of fixed points are connected and that the system can diffuse from one to the other. If the attracting and repelling lines were not connected, then all the fluctuation paths reaching a given point away from these lines would necessarily be instantons having a non-zero action at the $1/\nu$ order in the exponential of $P(A,B)$, which means that $P(A,B)$ would then globally satisfy an LDP with speed $1/\nu$. This is illustrated in a general way in Fig.~\ref{figgenarg1}. There we see that a point $x$ can be reached from an attracting set $A_s$ either directly with a single instanton (black path in Fig.~\ref{figgenarg1}a), or with a fluctuation path that reaches the repelling set $A_u$ before reaching $x$ (blue path). The second part of latter fluctuation path, going from $A_u$ to $x$, follows the natural dynamics of the system, and so carries zero action, while the first part, going from $A_s$ to $A_u$, is an instanton with non-zero action at the scale $1/\nu$. Consequently, the whole path must inherit the action of the instanton, which leads us to conclude that large deviations in this scenario are instanton-related and, thus, have the speed $1/\nu$.

If the sets $A_s$ and $A_u$ of fixed points are connected, as is the case for the $AB$ model, then the part of the fluctuation path that goes from the attracting set to the repelling set (see Fig.~\ref{figgenarg1}b) is not an instanton anymore, at least not an instanton with positive action at the scale $1/\nu$. In the $AB$ model, it is a simple diffusing path or \emph{sub-instanton} that contributes, as we have seen, to a factor $\E^{-c/\sqrt{\nu}}$ in the stationary probability density, where $c$ is some positive constant. In general, this is greater than the probability of ``true'' instantons, which scales like $\E^{-c/\nu}$ as $\nu\ra 0$. It might be the case that, for points close to the attracting set, and for small but finite values of $\nu$, instantons can have a probability greater than the probability of sub-instantons. This happens, for example, in the $AB$ model near the attracting line. However, the fact is that, because of their lower speed, sub-instantons necessarily have a probability higher than the probability of instantons in the limit $\nu\ra 0$, which means that they determine the dominant LDP.

We expect this scenario to apply to more realistic systems having extended and connected sets of steady states. In general, the probability of diffusing paths or sub-instantons going from an attracting set to a repelling set is likely to depend on the particular system considered, but what should be clear is that these paths have, in the low-noise limit, a probability greater than the probability of instantons. As a result, they must determine, if they exist, the dominant large deviation form of the stationary probability density of the system, when such a stationary density exists. In the future, it would be interesting to determine minimal conditions for the appearance of sub-instantons or related fluctuation paths, and to see whether these generally arise in stochastic systems violating detailed balance, as seems to be suggested by our results.

\section*{Acknowledgments}

We thank Eric Simonnet for first proposing to study the $AB$ model in relation to averaging techniques. We also thank Oleg Zaboronsky and Jason Laurie for interesting discussions. H.T. acknowledges the hospitality of the Laboratoire de physique at ENS Lyon, the LPTMC at the Universit\'e Paris 6, and the National Institute of Theoretical Physics at the University of Stellenbosch, where parts of this paper were written. The work of F.B. was supported by the ANR programs STATOCEAN (ANR-09-SYSC-014) and ANR STOSYMAP (ANR-2011-BS01-015). The work of H.T. was supported by an Interdisciplinary Research Fellowship (RCUK).

\section*{References}

\bibliography{ABmodelpaper}
\bibliographystyle{elsarticle-num}
\end{document}